\documentclass[10pt,conference]{IEEEtran}
\IEEEoverridecommandlockouts
\usepackage{cite}
\usepackage{amsmath,amssymb,amsfonts}
\usepackage{graphicx}
\usepackage{textcomp}
\usepackage{tabularx}
\usepackage{multirow}
\usepackage{lipsum}  
\usepackage[table,xcdraw]{xcolor}
\usepackage{hyperref} 
\usepackage{cleveref}
\usepackage{algpseudocode}
\usepackage{tcolorbox}
\usepackage{dialogue}
\usepackage{algorithm}
\usepackage{soul}

\newfloat{excerpt}{h}{lob}
\floatname{excerpt}{Excerpt}
\Crefname{excerpt}{Excerpt}{Excerpts}
\newcolumntype{Y}{>{\centering\arraybackslash}X}

\def\BibTeX{{\rm B\kern-.05em{\sc i\kern-.025em b}\kern-.08em
    T\kern-.1667em\lower.7ex\hbox{E}\kern-.125emX}}
\begin{document}

\title{APIDA-Chat: Structured Synthesis of API Search Dialogues to Bootstrap Conversational Agents}

\author{\IEEEauthorblockN{Zachary Eberhart and Collin McMillan}
\IEEEauthorblockA{\textit{Department of Computer Science} \\
\textit{University of Notre Dame}\\
Notre Dame, IN, USA \\
\{zeberhar, cmc\}@nd.edu}
}

\maketitle

\begin{abstract}
Large‑language‑model assistants are suitable for explaining popular APIs, yet they falter on niche or proprietary libraries because the multi‑turn dialogue data needed for fine‑tuning are scarce. We present \emph{APIDA‑Chat}, an open‑source pipeline that converts symbolic dialogue‑act ``scripts'' into realistic, domain‑grounded API Search conversations using a lightweight model for inexpensive training data generation. Phase I pairs a legacy dialogue planner with a high‑capability teacher LLM (o4‑mini) to synthesize a ``gold set'' of realized dialogues; then, a smaller Llama 3.2 3B student model is fine-tuned on this corpus. Phase II drops the teacher and reuses the same planner with the fine-tuned model, allowing rapid, low‑cost synthesis of new dialogues without exposing source code to external services. The fine‑tuned student improves BLEU from 0.38 to 0.50 and BERTScore from 0.88 to 0.91 versus the base model while running entirely on a single consumer GPU. All components are modular and publicly released to serve as a conservative baseline for future work. \emph{APIDA-Chat} is open-sourced at \url{https://github.com/Zeberhart/apida-chat} and a video demo is available at \url{https://youtu.be/YqmZBHyGbPs}.
\end{abstract}


\section{Introduction}
Conversational assistants are rapidly becoming a standard interface for programmers who need to search APIs, debug code snippets, or explore unfamiliar libraries. A recent StackOverflow survey demonstrated that over 60\% of developers already rely on conversational AI tools, and over 70\% of those are currently using or interested in using those tools to learn about a codebase~\cite{StackOverflow2024Survey}; qualitative studies echo these findings, identifying API exploration as a top use-case for AI assistants~\cite{Sergeyuk_2025}. These chatbots rely on large language models (LLMs) that are trained on corpora of multi-turn dialogues describing real programming tasks.

Yet obtaining such \emph{domain-specific} dialogue data remains difficult. Public corpora are rich in data pertaining to popular ecosystems, but lack data about niche or private codebases. \emph{Retrieval-augmented generation} (RAG) can help LLMs overcome this knowledge gap, but conversational data is still important to shape the behavior of a model: style, safety rules, and how to integrate retrieved snippets. Fine-tuning a model on relevant training data can also increase accuracy in retrieval tasks beyond RAG alone, particularly in smaller models~\cite{Soudani_2024}. 

Therefore, training an assistant to specialize in an unfamiliar domain requires a method for collecting domain-specific training dialogues. Creating fresh conversations by hiring experts is slow and expensive, so companies and researchers increasingly turn to \emph{synthetic data} generation: prompting an LLM to talk to itself, then fine-tuning another model on the results. The \emph{Self-Instruct} framework popularized this idea for general instruction following~\cite{wang2023selfinstructaligninglanguagemodels}, and recent surveys list dozens of LLM-generated datasets across text and code~\cite{Nad__2025}.

However, naïvely sampling ``self-chats'' suffers from a number of short-comings. One concern is \emph{coverage complexity}: artificially-generated samples should be made to avoid drifting towards ``happy-path'' scenarios and ensure domain coverage by visiting edge cases such as ambiguous queries or repeated failures\cite{cui2024adainstructadaptinginstructiongenerators, steindl-etal-2025-coprus}. Another challenge is \emph{domain grounding}, as LLMs frequently hallucinate API elements or usage details that can corrupt downstream training~\cite{liu2024exploring}. Additionally, free-form generations offer no latent structure for debugging or content-filtering, leading to poor \emph{audibility}~\cite{wei2023chainofthoughtpromptingelicitsreasoning}. 

Beyond issues of output-quality, organizations often face strict privacy constraints that rule out sending proprietary code to external AI services, and the price of invoking frontier-grade models quickly becomes prohibitive when scaling to thousands or millions of synthetic dialogues. To enable efficient development of conversational assistants for software engineering tasks, we seek an \emph{efficient} data-generation method that is controllable, reliable, and auditable. 

\begin{figure*}[t!]
    \centering
    \includegraphics[width=1\textwidth]{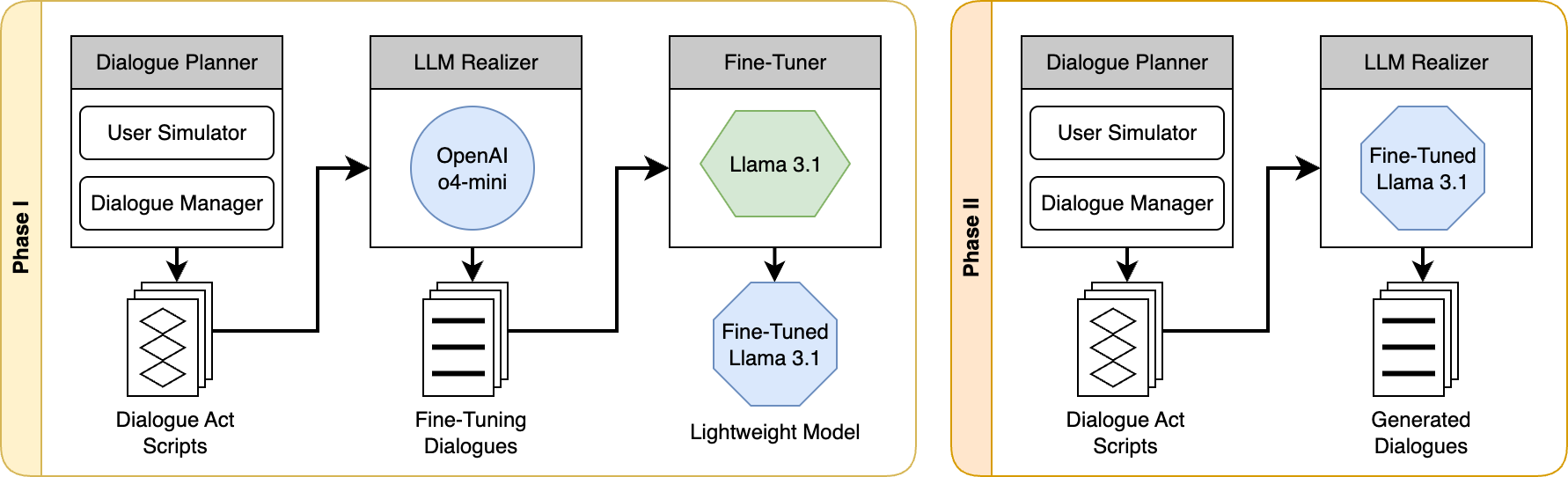}
    \caption{Two-phase pipelines.  Phase I bootstraps a lightweight model; Phase II reuses the planner with the fine-tuned model for low-cost dialogue generation.}
    \label{fig:pipeline}
\end{figure*}

In this paper, we present a pipeline to generate \emph{structured} synthetic conversational data for the API search domain. We repurpose a \emph{dialogue manager} (DM) for interactive API search as a \emph{planner}. The DM emits a conversational ``script'' grounded in a target API comprising a sequence of user- and system- dialogue acts (DAs), e.g., \texttt{Provide-Query}, \texttt{Suggest}, \texttt{Elicit-Info}. A modern LLM realizes each act as a fluent, natural-language utterance, and a lightweight model is then distilled from the synthetic chats. The planner guarantees structure; the LLM supplies natural language. By decoupling symbolic planning from neural realization, the pipeline mass-produces domain-grounded dialogues at low cost while exposing every step as an inspectable trace.

We demonstrate this pipeline in a tool, \emph{APIDA-Chat} (API Dialogue Act Chat), using publicly available services and data. There are 3 components, as illustrated in ~\Cref{fig:pipeline}: 
\begin{enumerate}
    \item  \textbf{Dialogue Planner} -- a 2021 DM for API search~\cite{eberhart2021dialoguemanagementinteractiveapi} emits dialogue act scripts referencing real API symbols.
    \item  \textbf{LLM Realizer} -- a high-quality LLM consumes a script and produces a multi-turn chat realizing every act.
    \item \textbf{Fine-Tuner} -- a LoRA adapter trains a lightweight model on a small corpus of high-quality synthetic dialogues.
\end{enumerate}
The fine-tuned model can then be used as the LLM realizer in conjunction with the dialogue planner to generate synthetic dialogues on controlled hardware, at a fraction of the cost. 

Our reference implementation relies on deliberately lightweight components, but guarantees act-level coverage and domain grounding that naïve self-chat lacks. Stronger planners or retrieval-augmented generators are drop-in replacements, so our demonstration establishes a conservative lower bound.

We contribute: (1) a \textbf{planner-guided data generator} for bootstrapping API search chatbots; (2) an \textbf{open corpus and code release}, including 300 script-chat pairs for the Allegro C API; and (3) early \textbf{empirical evidence} that our fine-tuning pipeline enables a lightweight model to generate dialogues comparable to those produced by an advanced-reasoning model. This tool aims to serve  researchers and enterprise teams looking to generate conversational training data for niche or private APIs, as well as those with rule-based API search tools who want to integrate LLM fluidity.

\section{Pipeline Overview}

The pipeline in Figure~\ref{fig:pipeline} shows how \emph{APIDA-Chat} turns a handful of API-aware “scripts” into an unlimited supply of domain-grounded training data. \emph{Phase I} bootstraps a compact model using ``teacher-student'' pattern, and \emph{Phase II} drops the costly teacher, using the fine-tuned student model to generate new dialogues at a fraction of the compute cost. Each component in our open-source tool is modular, and can be swapped for stronger/customized tools without modifying the rest of the pipeline. 

\subsection{Phase I -- Bootstrapping a Fine-Tuned Model}
In Phrase I, a dialogue planner writes a symbolic script, a frontier-grade ``teacher'' LLM realizes the scripts into natural language conversations, and a LoRA fine-tuner distills those dialogues into a lightweight ``student'' model.

\subsubsection{Dialogue Planner}
The purpose of the dialogue planner is to generate a sequence of dialogue acts that might be observed in a real conversation in which the assistant efficiently guides the user to their goal. A dialogue act (DA) is a high-level abstraction of a conversational turn comprising a dialogue act type and relevant variables -- e.g., \texttt{Provide-Query("bitmaps four needed dynamically")}. By emitting acts rather than surface text, the planner helps guarantee intent coverage, enforce safety rules, and determine concrete API entities for downstream components.

We use a dialogue planner for a conversational API search system published by Eberhart et al.~\cite{eberhart2021dialoguemanagementinteractiveapi}. The planner is split into a \emph{user simulator} that follows heuristic rules derived from a Wizard-of-Oz study and a \emph{dialogue manager} trained via deep Q-learning. The key idea is to model a user who exhibits realistic behavior, and an assistant that follows an optimal policy to help the user with their task. At each turn, the simulator chooses a user act based on conversation state and stochastic behavior parameters.  The DM observes that act, queries a TF–IDF retriever over a knowledge base of API documentation, and outputs a system act. We demonstrate the tool with a dialogue policy trained using 5 million self-play steps and a turn-penalty reward (details unchanged from \cite{eberhart2021dialoguemanagementinteractiveapi}).

\subsubsection{LLM Realizer}
Next, we use an advanced-reasoning LLM to generate a small, high-quality dataset of natural-language dialogues. These dialogues will be used to fine-tune our final model, so it is important to use an LLM capable of generating realistically expressive and varied search scenarios and dialogue turns. We adopt OpenAI’s \texttt{o4-mini} (Apr 2025 snapshot) as the teacher LLM because it delivers strong code-reasoning and fluent dialogue at a reasonable cost.

We embed the full DA script in a single prompt and ask o4-mini to produce the whole dialogue in one pass. One-shot realization is roughly six times lighter on context tokens than alternating user/assistant calls, and it can yield more coherent exchanges because the model can plan across turns. Prior work on shopping agents~\cite{li2025wizardshoppingtargetorientedecommerce} and interview generation~\cite{debaer2025singlevsdualpromptdialogue} report mixed results comparing one-shot and multi-step prompting, so we adopt the cheaper option by default.

The prompt itself has two blocks. The \emph{system} message frames the task and packs several practical rules -- invent a plausible project context, weave every supplied keyword into realistic code or prose, open with a bridge sentence when the next act diverges from the user’s last remark, vary tone and length on both sides, keep cross-turn references coherent, and never reveal API symbols unless the script calls for them. The subsequent \emph{user} message embeds concise DA definitions, reiterates style constraints, and finally appends the DA script itself. The model must return a single JSON array of realized turns. These instructions were refined over multiple ad-hoc prompt iterations; they live in a separate text file so practitioners can adjust wording or add domain-specific constraints without touching code.

\subsubsection{Fine-Tuner}
This step distills the teacher-generated conversations into a compact student model that fits on a single consumer GPU. We start from a \texttt{Llama-3.2-3B-Instruct} model: an instruction-tuned model whose 4-bit quantization keeps memory demands low. We fine-tune the model by training a lightweight LoRA adapter over the synthetic corpus with a modest learning rate, leaving the original weights unchanged.

\subsection{Phase II -- Production-Scale Dialogue Synthesis}
In Phase II, the same dialogue planner produces fresh DA scripts, but now the fine-tuned student realizes each script locally. Users may loop this phase indefinitely, schedule nightly runs on new APIs, or plug the student into an online system: the pipeline’s earlier stages remain unchanged, and higher-capacity planners or retrieval components can be swapped in.

\subsection{Limitations}
The current prototype inherits several constraints from its lightweight design. First, the legacy dialogue planner used was trained on a single pre-processed API with a TF-IDF retriever, so its action space and grounding quality are narrow and fairly brittle. Second, the teacher realizer answers from parametric knowledge only, and is not immune to hallucination. Third, our prompt template is hand-tuned and may need adaptation in other domains, while the seed set of 250 dialogues risks style bias and gaps in edge-case coverage. Finally, the pipeline still depends on a proprietary cloud model in Phase I, which some privacy-conscious organizations cannot use.

These weaknesses are largely modular. A newer or custom planner can drop into the existing interface, and a retrieval-augmented realizer would help curb hallucinations. Because prompts live in a standalone text file, prompt tuning is straightforward. Phase I can be looped to enlarge the synthetic corpus, optionally mixing in selective human review. Lastly, teams that require full on-prem control can switch the teacher to an open 70 B model. Taken together, these upgrade paths suggest that our results should be viewed as a conservative baseline rather than an upper bound.

\section{Synthetic Dialogue Dataset}

\begin{table}[!t]
    
    \caption{Dialogue length, diversity, and similarity to teacher.}
    
\renewcommand{\arraystretch}{1.5}
\centering
\begin{tabular}{|c|cccc|}
\hline
\rowcolor[HTML]{C0C0C0} 
Model & Avg. Len & \# Unique & BLEU & BERTScore                     \\ \hline

\cellcolor[HTML]{EFEFEF}\begin{tabular}[c]{@{}c@{}} o4-mini \end{tabular} & 99.96   & 6925 & N/A & N/A                      \\\hline

\renewcommand{\arraystretch}{1}
\cellcolor[HTML]{EFEFEF}\begin{tabular}[c]{@{}c@{}} llama \end{tabular}   & 82.58     & 3147   & .381 & .881 \\
\hline

\cellcolor[HTML]{EFEFEF}\begin{tabular}[c]{@{}c@{}} llama-ft \end{tabular}   & 104.88     & 5138   & .497  & .906 \\
\hline

\end{tabular}
    \vspace{-.4cm}
    \label{tab:dataset}
\end{table}

To illustrate the full pipeline, we demonstrate the \emph{APIDA-Chat} tool targeting the \texttt{Allegro 5.0} C multimedia library -- the same domain used in the original dialogue-manager study. Phase I automatically generates \textbf{250} DA scripts, realizes them with o4-mini, and fine-tunes the Llama-3.2 3B student. Phase II then produces an additional \textbf{50} held-out scripts that are realized by all three models (teacher, base student, fine-tuned student) for comparison.

\subsection{Dataset}
The resulting training set contains 250 multi-turn conversations (\(\approx\)180 k tokens).  \Cref{tab:dataset} summarises basic statistics: the fine-tuned student matches the teacher’s average turn length (105 vs 100 tokens) and nearly doubles the vocabulary size of the untuned student, indicating richer lexical diversity.  Against the teacher references, BLEU rises from 0.38 to 0.50 and BERTScore from 0.88 to 0.91 after fine-tuning.  These intrinsic scores measure similarity to the teacher, not end-user quality; a fuller evaluation with human or automated preference judgments is left to future work.

\subsection{Examples}
\begin{figure}[b!]
    \centering
    \label{exc:followup}
\begin{tcolorbox}[left=0pt,right=-25pt,top=1pt,bottom=1pt]
\begin{flushleft}

\texttt{Turn 1 (User)}
\newline
\newline
 \begin{tabularx}{.9\columnwidth}{lX}
    \textbf{DA}  & \emph{Provide-Query("distinct 5000 holes fixasin")}\\
    \rowcolor[HTML]{D0D0D0} 
    \textbf{o4-mini}  & I’m trying to triangulate a shape with 5000 distinct holes, but my custom fixasin logic keeps failing—any ideas?\\
    \textbf{llama}  & I'm trying to create a 5000-hole fixasin, but I'm not sure what I'm doing—can you help me out?\\
    \rowcolor[HTML]{D0D0D0} 
    \textbf{llama-ft}  & I’m trying to fixasin a 5000 hole polygon in my 3D model—any ideas on how to handle that distinct geometry?\\
    \hline
\end{tabularx}
\newline
\newline
\texttt{Turn 2 (System)}
\newline
\newline
 \begin{tabularx}{.9\columnwidth}{lX}
    \textbf{DA}  & \emph{Suggest(al\_fixasin)}\\
    \rowcolor[HTML]{D0D0D0} 
    \textbf{o4-mini}  & You might try al\_fixasin(); it computes a fixed-point arcsin robustly, which could stabilize your angle math in those hole vertices.\\
    \textbf{llama}  & Have you considered using `al\_fixasin`? It's a function that can help you fixasin a given number of holes.\\
    \rowcolor[HTML]{D0D0D0} 
    \textbf{llama-ft}  & You might try al\_fixasin() to convert your angle to a fixed-point representation, which can help with precision in your 3D math.\\
    \hline
\end{tabularx}
\newline
\newline
\texttt{Turn 3 (User)}
\newline
\newline
 \begin{tabularx}{.9\columnwidth}{lX}
    \textbf{DA}  & \emph{Elicit-Info(al\_fixasin)}\\
    \rowcolor[HTML]{D0D0D0} 
    
    \textbf{o4-mini}  & Can you give me the details on al\_fixasin?\\
    \textbf{llama}  & What exactly does `al\_fixasin` do?\\
    \rowcolor[HTML]{D0D0D0} 
    
    \textbf{llama-ft}  & Can you give me the details on al\_fixasin? What input and output does it take?\\
    \hline
\end{tabularx}
\newline
\newline
\texttt{Turn 4 (System)}
\newline
\newline
 \begin{tabularx}{.9\columnwidth}{lX}
    \textbf{DA}  & \emph{Info(al\_fixasin)}\\
    \rowcolor[HTML]{D0D0D0} 
    
    \textbf{o4-mini}  & al\_fixasin(AL\_FIXED x) returns sin\textsuperscript{-1}(x) as a 16.16 fixed-point value; it clamps inputs outside [-1,1] and avoids floating-point.\\
    \textbf{llama}  & `al\_fixasin` is a function that takes an integer as input and returns a fixed-point value. It's often used in graphics and game development to fixasin holes in a shape.\\
    \rowcolor[HTML]{D0D0D0} 
    
    \textbf{llama-ft}  & al\_fixasin(AL\_FIXED y) returns the inverse sine of y in fixed-point format (16.16). It’s zero-alloc and safe to use in tight loops.\\
\end{tabularx}
\end{flushleft}
\end{tcolorbox}
\label{exc:followup}
    \caption{Example turns generated by different models for a given DA script.}
    \label{fig:example}
\end{figure}

To give a small glimpse into the behavior of the three models, \Cref{fig:example} juxtaposes the first four turns of the \emph{same} dialogue-act script realized by each model. In the first turn, the user provides a query; in the second, the AI assistant suggests a relevant function; in the third, the user requests information about the function; and in the fourth, the assistant attempts to answer the user's question.

The teacher model (o4-mini) transforms the sparse keyword list -- ``distinct, 5000, holes, fixasin’’ -- into a plausible scenario (“triangulate a shape with 5000 distinct holes...”) and then weaves that scenario into subsequent turns. The fine-tuned student (llama-ft) mirrors this structure but betrays its smaller capacity: it misinterprets fixasin as a verb (``fixasin a 5000 hole polygon''), yet still respects the act sequence and maintains cross-turn coherence. The baseline model (llama) illustrates the importance of the synthetic corpus: it drops keywords, gives a generic suggestion that ignores the user’s context, and offers a vague description of the \texttt{al\_fixasin} function that lacks salient details (signature, range, allocation safety) provided by the other two models.

In short, fine-tuning on just a few hundred planner-guided dialogues closes much of the gap to the teacher, while the zero-shot model frequently falls short on both grounding and conversational flow.

\subsection{Reproduction \& Access}
\emph{APIDA-Chat} is publicly available under at \url{https://github.com/Zeberhart/apida-chat}. The repository bundles command-line scripts for each pipeline component together with a notebook that invokes them sequentially. Running the notebook on a standard Colab T4 instance reproduces all artifacts reported in this paper without code modification.
\section{Conclusion}

\emph{APIDA-Chat} demonstrates that a lightweight dialogue planner, a high-capability teacher LLM, and a modest LoRA fine-tune are sufficient to bootstrap a domain-grounded conversational model for API search. By separating symbolic planning from realization, the pipeline delivers three practical advantages: (1) DA-level coverage and grounding, (2) inspectable generation traces for safety and debugging, and (3) a compact student that can run locally without per-call cloud fees.

The architecture is intentionally modular; stronger planners, retrieval-augmented realizers, or alternative teacher/student models can be swapped in with minimal effort. Future work will explore improved RAG, automatic prompt search, and human-in-the-loop filtering to further improve coverage and factual accuracy. We hope the tool and dataset released with this paper will serve as a reproducible starting point for specialized, controllable dialogue data generation in the software-engineering domain.

\bibliographystyle{IEEEtran}
\bibliography{main}

\end{document}